# Magnetization Process and Adiabatic Demagnetization of the Antiferromagnetic Spin-1/2 Heisenberg Cubic Cluster

J. STREČKA*, J. ČISÁROVÁ

Institute of Physics, Faculty of Science, P. J. Šafárik University, Park Angelinum 9, 040 01 Košice, Slovakia

A full energy spectrum of the spin-1/2 Heisenberg cubic cluster is used to investigate a low-temperature magnetization process and adiabatic demagnetization of this zero-dimensional 2x2x2 quantum spin system. It is shown that the antiferromagnetic spin-1/2 Heisenberg cube exhibits at low enough temperatures a stepwise magnetization curve with four intermediate plateaux at zero, one quarter, one half, and three quarters of the saturation magnetization. We have also found the enhanced magnetocaloric effect close to level-crossing fields that determine transitions between the intermediate plateaux.

## 1. Introduction

Molecular nanomagnets represent a hot topic of current research interest, which has been stimulated mainly by an immense application potential of single-molecule magnets in a development of new generation of high-density storage devices [1]. Beside this, small magnetic molecules may exhibit other fascinating macroscopic quantum phenomena such as the quantum tunnelling of magnetization [1] or the existence of striking plateaux and jumps in low-temperature magnetization curves [2]. This latter quantum effect is usually accompanied in the most Heisenberg spin clusters with the enhanced magnetocaloric effect [2]. It is noteworthy that magnetic properties of a few Heisenberg spin clusters of basic geometry still remain unexplored, yet.

Although a magnetic behavior of the spin-1/2 Heisenberg cubic cluster has been examined to a certain extent in Refs. [3-5], it surprisingly turns out that the magnetization process at finite temperatures and the magnetocaloric effect have not been dealt with previously. In addition, we also aim to provide exact results for a full energy spectrum of the spin-1/2 Heisenberg cube, because the results reported on previously in Refs. [3-5] are in discrepancy for some particular energy eigenvalues.

## 2. Heisenberg cubic cluster

Let us consider the spin-1/2 Heisenberg model defined on the 2x2x2 cubic cluster through the Hamiltonian

$$\hat{H} = J\,(\hat{S}_1\cdot\hat{S}_2 + \hat{S}_2\cdot\hat{S}_3 + \hat{S}_3\cdot\hat{S}_4 + \hat{S}_4\cdot\hat{S}_1 + \hat{S}_5\cdot\hat{S}_6 + \hat{S}_6\cdot\hat{S}_7 + \hat{S}_7\cdot\hat{S}_8 + \hat{S}_8\cdot\hat{S}_5 + \hat{S}_1\cdot\hat{S}_5 + \hat{S}_2\cdot\hat{S}_6 + \hat{S}_3\cdot\hat{S}_7 + \hat{S}_4\cdot\hat{S}_8) - h\sum_{i=1}^{8}\hat{S}_i^z, \qquad (1)$$

where $\hat{S}_i \equiv (\hat{S}_i^x, \hat{S}_i^y, \hat{S}_i^z)$ denotes the standard spin-1/2 operator and the subscript $i$ specifies lattice sites according to labelling scheme shown in Fig.1. The last Zeeman's term from Eq. (1) commutes with the total Hamiltonian and hence, it is sufficient to find a full energy spectrum of the spin-1/2 Heisenberg cubic cluster in a zero field following the group-theoretical calculations developed in Refs. [3-5]. All eigenvalues of the spin-1/2 Heisenberg cube in a zero field are explicitly quoted in Tab.1 along with their respective degeneracies. The relevant eigenvalues of the Hamiltonian (1) then readily follow from the relation

$$E_j\,(h) = E_j\,(h=0) - h\,S_T^z\,, \qquad (2)$$

where $S_T^z = -S_T\,, -S_T+1, \ldots, S_T$ determines *z*-component of the total spin of the 2x2x2 cubic spin cluster.

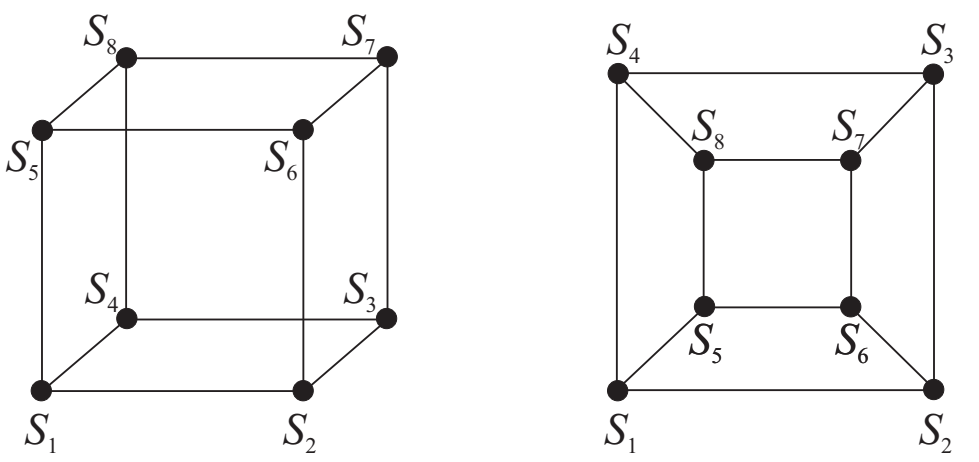

Fig.1. Two graphical representations of a cubic spin cluster.

Tab.1. Energy eigenvalues of the spin-1/2 Heisenberg cube in a zero magnetic field sorted according to the total spin $S_T$.

| $S_T$ | Energy [$J$] | Degeneracy |
|---|---|---|
| 4 | 3 | 1 |
| 3 | 0 | 1 |
| 3 | 1 | 3 |
| 3 | 2 | 3 |
| 2 | $-1 \pm 2^{1/2}$ | 1 |
| 2 | $\pm 1$ | 6 |
| 2 | $(1 \pm 5^{1/2}) / 2$ | 2 |
| 2 | 0 | 2 |
| 1 | -4 | 1 |
| 1 | -1 | 2 |
| 1 | 0 | 4 |
| 1 | $(-1 \pm 5^{1/2}) / 2$ | 3 |
| 1 | $(-3 \pm 5^{1/2}) / 2$ | 3 |
| 1 | $-2/3 + 400^{1/2} \cos(\alpha_1 + n2\pi/3)/3$ | 3 |
| | $\alpha_1 = \arccos(1000^{-1/2})/3$, $n$=0,1,2 | |
| 0 | $-1 \pm 2^{1/2}$ | 2 |
| 0 | -1 | 3 |
| 0 | -2 | 3 |
| 0 | 0 | 1 |
| 0 | $-5/3 - 112^{1/2} \cos(\alpha_2 + n2\pi/3)/3$ | 1 |
| | $\alpha_2 = \arccos(13/5488^{1/2})/3$, $n$=0,1,2 | |

*corresponding author; e-mail: jozef.strecka@upjs.sk

The free energy of the spin-1/2 Heisenberg cube can be obtained from the eigenvalues listed in Tab.1 and Eq. (2)

$$F = -k_{\mathrm{B}}\,T \ln \sum_j \exp\left(-E_j / k_{\mathrm{B}}\,T\right). \qquad (3)$$

Other thermodynamic quantities, such as the magnetization or entropy, can be now straightforwardly calculated from the free energy (3) using the basic relations of thermodynamics.

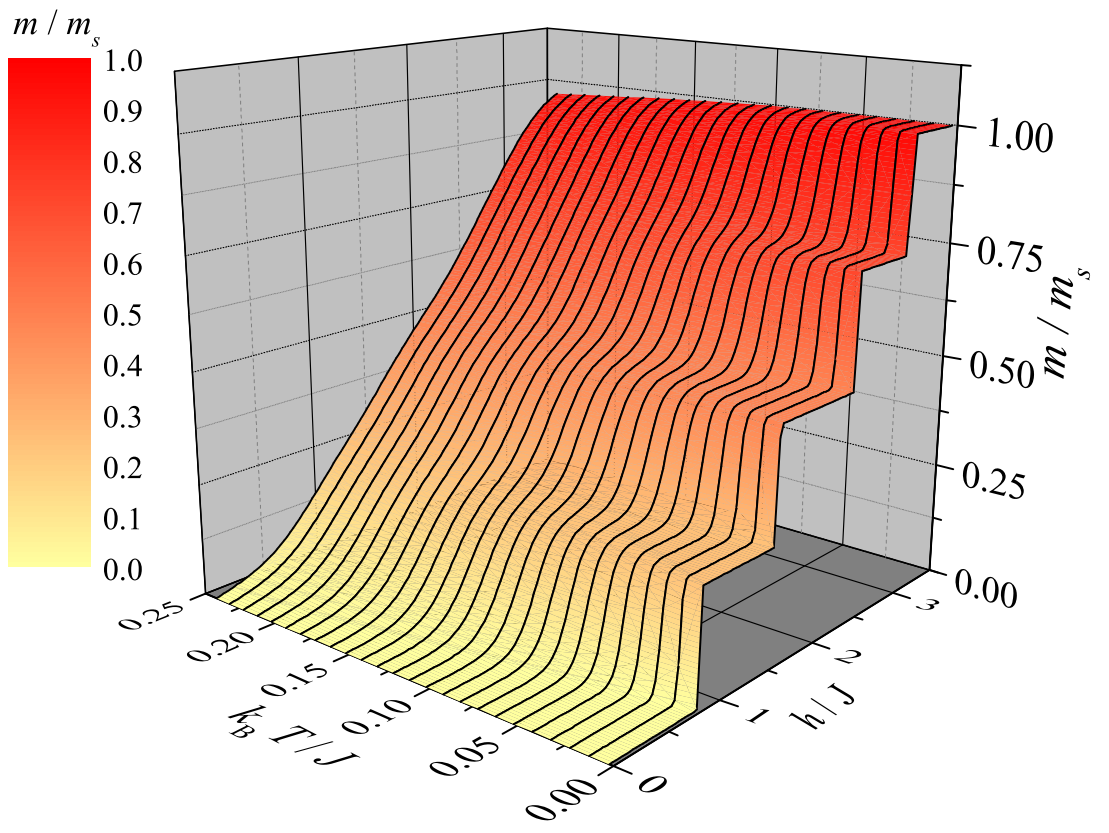

Fig.2. 3D surface plot of the magnetization (normalized with respect to its saturation value) as a function of the external magnetic field and temperature.

### 3. Results and Discussion

Let us proceed to a discussion of the most interesting results for the magnetization process and adiabatic demagnetization. The magnetization normalized with respect to its saturation value is plotted in Fig.2 against the magnetic field and temperature. The antiferromagnetic spin-1/2 Heisenberg cube exhibits at zero temperature a stepwise magnetization curve with four intermediate plateaux at zero, one quarter, one half, and three quarters of the saturation magnetization, whereas all four level-crossing fields $h_c / J$ = 0.82, 1.59, 2.41 and 3.00 assigned to the magnetization jumps can be easily found by equating the lowest-energy levels from the sectors with different $S_T$ (see Tab.1). It can be clearly seen from Fig.2 that the rising temperature gradually smoothens a stepwise magnetization curve until all magnetization plateaux and jumps completely disappear from the magnetization curve above $k_{\mathrm{B}}\,T / J \approx 0.20$.

Next, our particular attention is focused on the adiabatic demagnetization of the spin-1/2 Heisenberg cubic cluster. It is quite obvious from Fig.3 that the temperature shows a marked isentropic dependence on the magnetic field in a vicinity of the level-crossing fields, where an abrupt drop in temperature can be followed by a successive steep increase upon sweeping the magnetic field. An optimal value of the entropy for observing the enhanced magnetocaloric effect is $S = k_{\mathrm{B}}$ ln2, under which temperature goes to zero infinitely fast as the external field approaches level-crossing fields. However, it should be stressed that the entropy must be kept sufficiently close to this optimal value during the adiabatic demagnetization in order to warrant the efficient cooling (heating) just above (below) the level-crossing fields.

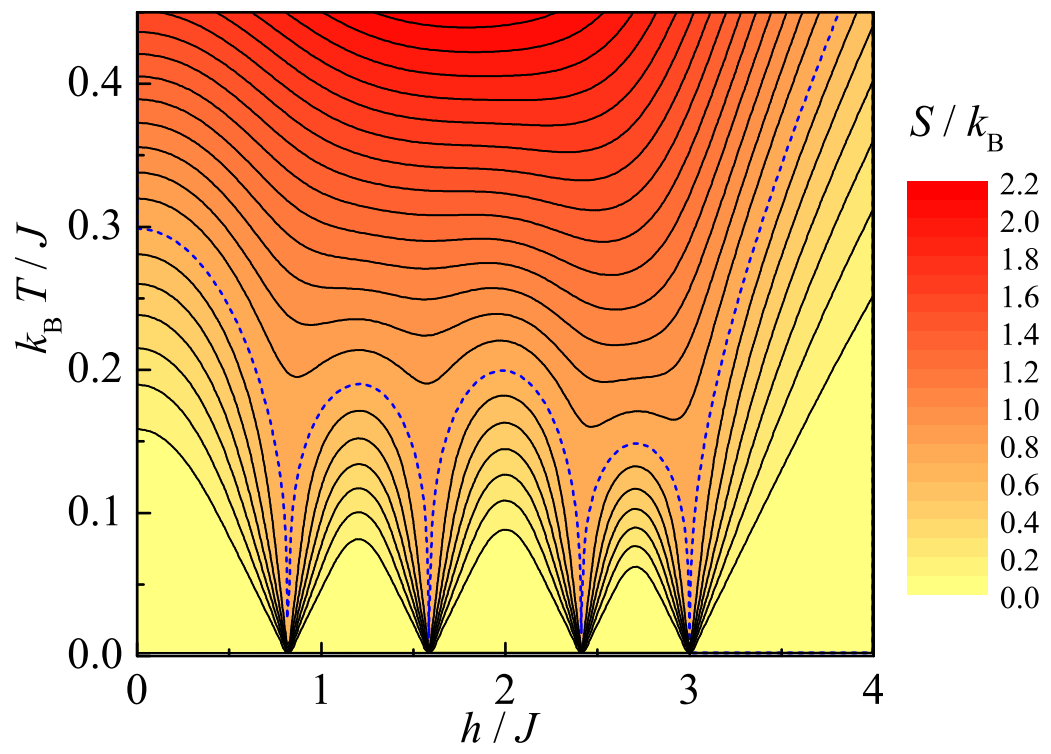

Fig.3. Adiabatic (isentropic) changes of temperature upon varying a magnetic field. A broken line shows the steepest variation of temperature with a magnetic field for the optimal value of entropy $S / k_{\mathrm{B}}$= ln 2 = 0.69315...

### 4. Conclusions

The present work is devoted to theoretical investigation of the magnetization process and magnetocaloric effect in the antiferromagnetic spin-1/2 Heisenberg cubic cluster. We have explored in particular how a marked magnetization curve with four intermediate plateaux at zero, one quarter, one half and three quarters of saturation magnetization is gradually smoothened due to thermal fluctuations, which are promoted by increasing temperature.

In addition, it has been demonstrated that the antiferromagnetic spin-1/2 Heisenberg cube exhibits an enhanced magnetocaloric effect close to level-crossing fields, which determine transitions between different magnetization plateaux. The most efficient cooling (heating) has been found for isentropic processes with the fixed value of entropy sufficiently close to the optimal value $S = k_{\mathrm{B}}$ ln 2.

### Acknowledgement

This work was financially supported by the projects VEGA 1/0234/12, APVV-0132-11, ITMS 26220120047.

### References

[1] J.R. Friedman, M.P. Sarachik, *Annu. Rev. Condens. Matter Phys.* **1**, 109 (2010).
[2] J. Schnack, *J. Low. Temp. Phys.* **142,** 279 (2006).
[3] G. Dresselhaus, *Phys. Rev.* **126**, 1664 (1962).
[4] Ch. Kawabata, M. Suzuki, *J. Phys. Soc. Jpn.* **28**, 16 (1970).
[5] P.N. Moustanis, S. Thanos, *Physica B* **202**, 65 (1994).